\documentclass[review]{elsarticle}
\usepackage{amsmath,amssymb}
\usepackage{graphicx}
\journal{Journal of \LaTeX\ Templates}

\begin{document}

\begin{frontmatter}

\title{Odd tensor E3 transitions and the generalized seniority in Sn-isotopes}

%\author[mymainaddress]{Bhoomika Maheshwari\corref{mycorrespondingauthor}}
\author{Bhoomika Maheshwari\corref{mycorrespondingauthor}}
%and Ashok Kumar Jain} 

%\author[mymainaddress]{and Ashok Kumar Jain}
%\author[mymainaddress]{Bhoomika Maheshwari}
%[mycorrespondingauthor]{Bhoomika Maheshwari} 
\cortext[mycorrespondingauthor]{Corresponding author}
\ead{bhoomika.physics@gmail.com}
%\author[mymainaddress]{and Ashok Kumar Jain}
\author{and Ashok Kumar Jain}
%\address[mymainaddress]{Department of Physics, Indian Institute of Technology, Roorkee-247667, India.}
\address{Department of Physics, Indian Institute of Technology, Roorkee-247667, India.}

\begin{abstract}
In our recent paper [Phys. Lett. B 753, (2016) 122], we have shown that both the odd and even tensor electric transition probabilities exhibit similar behavior within the generalized seniority scheme in a multi-j environment. This microscopic approach was used to show for the first time the occurrence of seniority isomers in the $ {13}^-$ isomers of Sn-isotopes, which decay by odd tensor $E1$ transition to the same seniority ($\Delta v = 0$) state. In this letter, we extend our studies to odd tensor $E3$ transitions connecting different seniority states ($\Delta v = 2$), and show for the first time that the generalized seniority scheme explains reasonably well the systematics of the $B(E3)$ values for the $(0^+ \rightarrow 3_1^-)$ transitions in the Sn-isotopes. Additionally, we support these results by seniority guided Large Scale Shell Model (LSSM) calculations. The generalized seniority results are able to single out the most crucial valence space required in the LSSM calculations.
\end{abstract}

\begin{keyword}
\texttt{$3_1^-$ states, Sn-isotopes, Generalized seniority, Odd-tensor $E3$ transitions}
%\sep \LaTeX\sep Elsevier \sep template
%\MSC[2010] 00-01\sep  99-00
\end{keyword}

\end{frontmatter}

%\linenumbers

\section{Introduction}

\begin{figure}
\includegraphics[width=13cm,height=11cm]{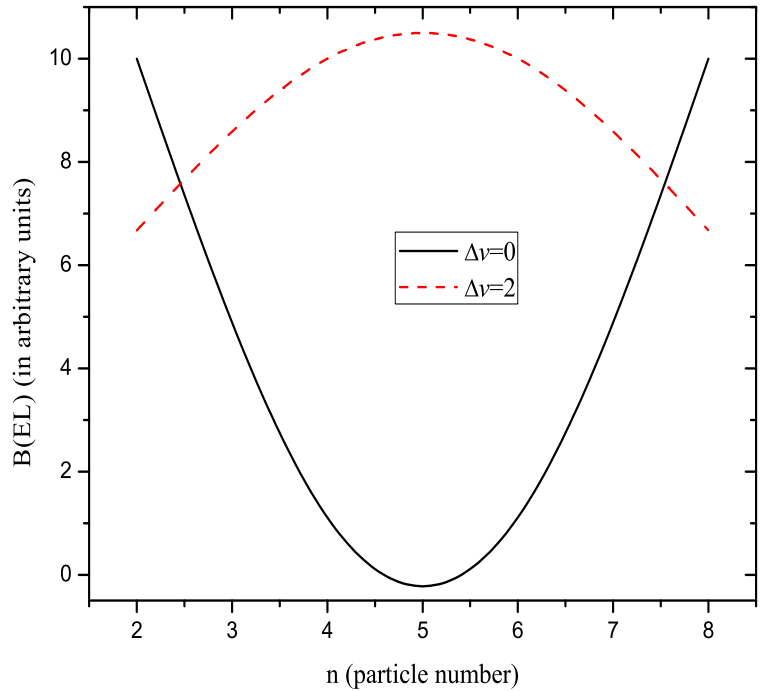} 
\caption{\label{fig:bel}(Color online) Schematic plot of the $B(EL)$ values, for both even and odd $L$, with particle number $n$, for seniority conserving $\Delta v=0$ transitions (solid line) and seniority changing $\Delta v=2$ transitions (dashed line), using a pair degeneracy of $\Omega=5$.} 
\end{figure}

The seniority scheme, first introduced by Racah~\cite{racah} in the atomic context, has been widely successful in explaining the spectroscopic features of semi-magic nuclei. A generalization of the single-j pure seniority scheme to multi-j generalized seniority was presented by Arima and Ichimura~\cite{arima}. Sn-isotopes represent the longest chain of isotopes in the nuclear chart. Because of the semi-magic nature of the Sn-isotopes, the seniority scheme ~\cite{racah, talmi, casten} has been extensively used to understand the various systematics in these nuclei ~\citep{morales, astier}. These also provide a rigorous testing ground for the various effective interactions used in the large scale shell model (LSSM) calculations ~\cite{simpson, maheshwari}. We have recently used the generalized seniority formalism for multi-j degenerate orbits to show that both the even and odd tensor electric transition probabilities for same seniority ($\Delta v = 0$) transitions follow similar trends, as schematically shown by the solid line in Fig.~\ref{fig:bel} ~\cite{maheshwari1}. Here, the reduced transition probabilities, $B(EL)$, follow a parabolic behavior with a minimum at the middle of the valence space. As a result, one expects larger half-life seniority isomers in the middle of the shell for both the even and the odd electric transitions. In our previous work ~\cite{maheshwari1}, we have convincingly shown that the $E1$, $\Delta v = 0$ transition probabilities from the ${13}^-$ isomers in Sn-isotopes behave similar to the $E2$, $\Delta v = 0$ transition probabilities of the ${10}^+$ and ${15}^-$ isomers in Sn-isotopes. As a consequence, the two classes of isomers may be treated on the same footing in this simple approach. It was also pointed out by us that the $B(EL)$ values for either odd or even $L$ in the case of seniority change by $\Delta v = 2$ exhibit an inverted parabolic behavior, as shown by the dashed line in Fig.~\ref{fig:bel}. In the present letter, we present the first evidence of the seniority changing $\Delta v = 2$ odd tensor transitions, where the $B(E3) \uparrow$ values for the $0^+$ to $3^-$ transitions in the Sn-isotopes, nicely fit into the inverted parabolic behavior obtained from our calculations.

To show this, we have used the $B(E3 \uparrow,0^+ \rightarrow 3_1^-)$ values of the first $3^-$ states in the Sn-isotopes from the compilation of Kibedi and Spear ~\cite{kibedi}, which decay to the $0^+$ ground state via $E3$ transition. These $3^-$ states are generally believed to be octupole vibrational in character. In 1981, Jonsson $et$ $al.$~\cite{jonsson} have reported the measurements of $3^-$ states in $^{112-124}$Sn using $(p,p' \gamma)$ reaction and Coulomb excitation, and also compared their measurements with previously known data and theoretical calculations, which were quite far from the experimental data. The behavior of the $3^-$ states, was later studied by Ansari and Ring ~\cite{ansari} within relativistic quasiparticle random-phase approximation (RQRPA), which highlighted the new challenges in fixing the force parameters for the $NL1$ and $NL3$ interactions to reproduce the experimental data. The RQRPA estimates were successful in obtaining the overall trend of the $B(E3)$ values. We now present a study of the $3^-$ states by using the simple microscopic approach based on generalized seniority scheme reported in our previous paper ~\cite{maheshwari1}. We find that the same scheme reproduces the systematics of $B(E3 \uparrow,0^+ \rightarrow 3_1^-)$ transition probabilities remarkably well. Remarkably, the involvement of the $d$ and $h$ orbits is required to explain the nature of $B(E3)$ values, which supports the previous interpretation of these states being octupole vibrational in character, as these orbits can be connected by a $\Delta l=3$ interaction. 

We have also carried out the seniority guided LSSM calculations by using the $d$ and $h$ orbits alone to validate our results, and are able to reproduce the measured systematics reasonably well. The generalized seniority scheme, therefore, turns out to be immensely useful in singling out the most critical valence space for the LSSM calculations. The $3^-$ states thus satisfy the generalized seniority scheme quite well. We present the details of the calculations and results in the next section.

\section{Calculations and results}

We have used the generalized seniority formalism for multi-j degenerate orbits presented in our recent paper~\cite{maheshwari1}, to calculate the $B(E3)$ values in the Sn-isotopes. The reduced $B(E3)$ transition probabilities for $n$ particles in the multi-j $\tilde{j}=j \otimes j' ....$ configuration can be obtained from the equation, 
\begin{eqnarray}
B(E3) \uparrow=\frac{7}{2J_i+1}|\langle \tilde{j}^n v l J_f || \sum_i r_i^3 Y^{3}(\theta_i,\phi_i) || \tilde{j}^n v' l' J_i \rangle |^2
\end{eqnarray}
where the reduced matrix elements between $\Delta v = 2$ states can be written in terms of seniority reduction formula as [5],
\begin{eqnarray}
\langle \tilde{j}^n v l J_f ||\sum_i r_i^3 Y^{3}|| \tilde{j}^n v\pm 2 l' J_i \rangle  = \Bigg[ \sqrt{\frac{(n-v+2)(2\Omega+2-n-v)}{4(\Omega+1-v)}} \Bigg] \nonumber\\ \langle \tilde{j}^v v l J_f ||\sum_i r_i^3 Y^{3}|| \tilde{j}^v v\pm 2 l' J_i \rangle 
\end{eqnarray}

The coefficients in the square brackets depend on the particle number $n$, the generalized seniority $v$ and the corresponding total pair degeneracy $\Omega= \frac{1}{2}(2 \tilde{j} +1)= \frac{1}{2} \sum \limits_j (2j+1)$ in the multi-j configuration. We have calculated the $B(E3) \uparrow$ values for the $3^-$ states in the Sn-isotopes by using two values of $\Omega= 9$ and $11$, corresponding to the $d_{5/2} \otimes h_{11/2}$, and $d_{5/2} \otimes d_{3/2} \otimes h_{11/2}$ valence spaces, respectively.

\begin{figure}
\includegraphics[width=13cm,height=11cm]{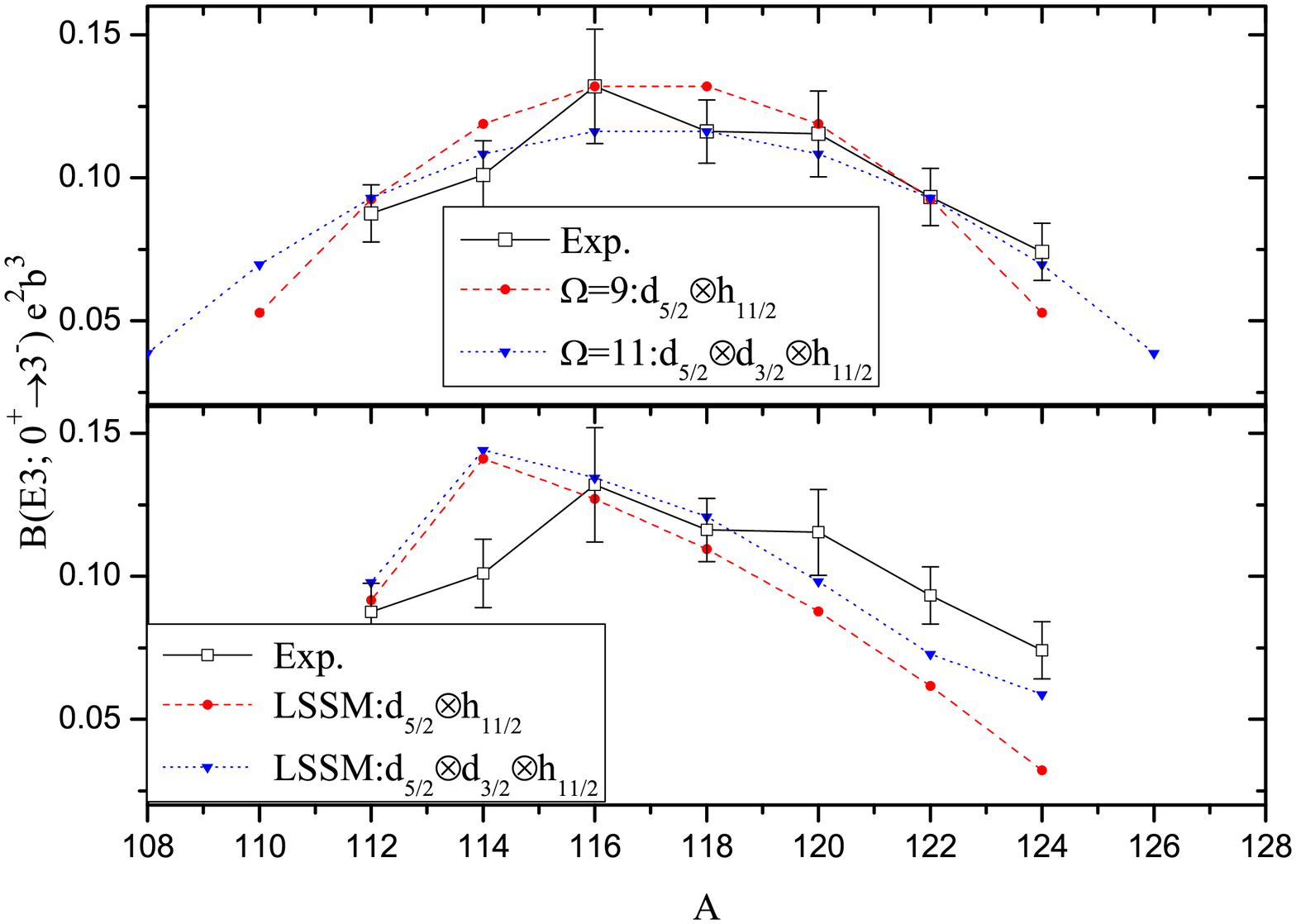}
\caption{\label{fig:be3}(Color online) A comparison of the experimental $B(E3)$ systematics in the Sn-isotopes (solid line) with the generalized seniority (upper panel) and LSSM calculations (lower panel). The dashed and dotted lines stand for a choice of the $d_{5/2} \otimes h_{11/2}$, and $d_{5/2} \otimes d_{3/2} \otimes h_{11/2}$ valence spaces, respectively.} 
\end{figure}

We calculate the complete systematics by fitting the $B(E3)$ values for $^{116}$Sn and $^{118}$Sn for $\Omega=9$ and $11$, respectively. We have taken $^{108}$Sn as the core for $\Omega=9$, while $^{106}$Sn as the core in the case of $\Omega=11$, where the core represents the $n=0$ situation. A comparison of the calculated results for both the $\Omega$ values with the known experimental data~\citep{kibedi} is shown in the upper panel of Fig.~\ref{fig:be3}. The measured values as well as the calculations show a peak in the middle of the given valence space. This behavior is as expected for odd tensor transitions taking place among the states having seniorities differing by $\Delta v=2$ (Fig.~\ref{fig:bel}). Hence, the simple generalized seniority scheme is able to explain the complete systematics with the valence space consisting of $d$ and $h$ orbits. We note that $\Omega=9$ corresponds to the $d_{5/2} \otimes h_{11/2}$ valence space and $\Omega=11$ corresponds to the $d_{5/2} \otimes d_{3/2} \otimes h_{11/2}$ valence space. The experimental data are better reproduced by the results for $\Omega=11$ suggesting that both the $d_{5/2}$ and $d_{3/2}$ orbits need to be included. As we show further the LSSM calculations carried out by us support this conclusion.

We have carried out the LSSM calculations by using the valence spaces of $d_{5/2} \otimes h_{11/2}$ and $d_{5/2} \otimes d_{3/2} \otimes h_{11/2}$, taking a cue from the generalized seniority calculations, though the active orbits, for the $50-82$ neutron valence space, are $0g_{7/2}$, $1d_{5/2}$, $1d_{3/2}$, $2s_{1/2}$, and $0h_{11/2}$. We have used the Nushell code~\cite{brown} and SN100PN interaction~\citep{brown1} for the LSSM calculations in the Sn-isotopes. The neutron single particle energies have been taken as -10.6089, -10.2893, -8.7167, -8.6944, -8.8152 MeV for the available $0g_{7/2}$, $1d_{5/2}$, $1d_{3/2}$, $2s_{1/2}$, and $0h_{11/2}$ valence orbits. We have taken the $g_{7/2}$ orbit as completely filled. The neutron effective charges of 1.5 and 1.6 have been used, for the $d_{5/2} \otimes h_{11/2}$ and $d_{5/2} \otimes d_{3/2} \otimes h_{11/2}$ valence spaces, respectively. The harmonic oscillator potential was chosen with an oscillator parameter of $\hbar \omega =45A^{-1/3}-25A^{-2/3}$. We have plotted the calculated results along with the experimental data in the lower panel of Fig.~\ref{fig:be3}, by dashed and dotted lines corresponding to the $d_{5/2} \otimes h_{11/2}$ and $d_{5/2} \otimes d_{3/2} \otimes h_{11/2}$ valence spaces. These seniority guided LSSM calculations reproduce the measured values quite well, except at $^{114}$Sn. The calculated results, for the $d_{5/2} \otimes d_{3/2} \otimes h_{11/2}$ valence space, reproduce the measured values much better. These calculations, hence, validate the choice of seniority guided valence space for the $3^-$ states in the Sn-isotopes. This procedure can be very useful to simplify the LSSM calculations, particularly where the dimensions become very large. However, the deviation of the shell model results for $^{114}$Sn isotope is very puzzling. This isotope has $N=64$, a possible magic number for octupole collectivity~\citep{nazarewicz, cottle}. Yet, the measured $B(E3)$ value for $^{114}$Sn is observed to be lower than $^{116}$Sn. Still, the RQRPA calculations of Ansari and Ring~\cite{ansari} do show a fall in the $B(E3)$ value for $^{114}$Sn as compared to $^{116}$Sn as seen in the experimental data. Our calculations by using the generalized seniority scheme also reproduce the results reasonably well. However, the LSSM calculations are not able to reproduce this fall in B(E3) for $^{114}$Sn.

\section{Conclusion}

To conclude, we have used the generalized seniority scheme for multi-j degenerate orbits to calculate the $B(E3 \uparrow,0^+ \rightarrow 3_1^-)$ transition probabilities for the $3^-$ states in the Sn-isotopes. Our calculations successfully reproduce the parabolic behavior of $B(E3)$ values with a peak in the middle due to the fact that the transition involves a seniority change of $2$. We also present LSSM calculations, using the $d$ and $h$ orbits from $50-82$ neutron valence space, as guided by the generalized seniority calculations. These shell model calculations reproduce the experimental data quite well, and support the generalized seniority scheme, except for a deviation at $^{114}$Sn. This explanation also supports the octupole vibrational character of the $3^-$ states, as they arise due to $d$ and $h$ orbits differing in $l$ values by $3$ units. However, the deviation of the LSSM results for $^{114}$Sn remains a puzzle. On the other hand, the RQRPA calculations as well as our calculations explain the overall trend pretty well. 

\section*{Acknowledgments}

Financial support from the Ministry of Human Resource Development (Government of India) is gratefully acknowledged.

%\section{Front matter}
%
%The author names and affiliations could be formatted in two ways:
%\begin{enumerate}[(1)]
%\item Group the authors per affiliation.
%\item Use footnotes to indicate the affiliations.
%\end{enumerate}
%See the front matter of this document for examples. You are recommended to conform your choice to the journal you are submitting to.
%
%\section{Bibliography styles}
%
%There are various bibliography styles available. You can select the style of your choice in the preamble of this document. These styles are Elsevier styles based on standard styles like Harvard and Vancouver. Please use Bib\TeX\ to generate your bibliography and include DOIs whenever available.

%Here are two sample references: \cite{Feynman1963118,Dirac1953888}.

%\section*{References}

%\bibliography{mybibfile}

\begin{thebibliography}{50}
\bibitem{racah}G. Racah, Phys. Rev. 63 (1943) 367.
\bibitem{arima}A. Arima, M. Ichimura, Prog. Theor. Phys. 36 (1966) 296.
\bibitem{talmi}I. Talmi, Simple Models of Complex Nuclei, (Harwood Academic, 1993).
\bibitem{casten}R. F. Casten, Nuclear Structure from a Simple Perspective, (Oxford University Press, 1990).
\bibitem{morales}I. O. Morales, P. Van Isacker, I. Talmi, Phys. Lett. B 703 (2011) 606. 
\bibitem{astier}A. Astier $\it{et}$ $\it{al.}$, Phys. Rev. C 85 (2012) 054316.
\bibitem{simpson}G. S. Simpson $\it{et}$ $\it{al.}$, Phys. Rev. Lett. 113 (2014) 132502.
\bibitem{maheshwari}B. Maheshwari, A. K. Jain and P. C. Srivastava, Phys. Rev. C 91 (2015) 024321, and references therein.
\bibitem{maheshwari1}B. Maheshwari, A. K. Jain, Phys. Lett. B 753 (2016) 122.
\bibitem{kibedi}T. Kibedi, and R. H. Spear, Atomic Data and Nuclear Data Tables 80 (2002) 35.
\bibitem{ansari}A. Ansari, and P. Ring, Phys. Rev. C 74 (2006) 054313.
\bibitem{jonsson}N. G. Jonsson et al., Nucl. Phys. A 371 (1981) 333.
\bibitem{brown}B. A Brown and W. D. M. Rae, Nushell@MSU, MSU-NSCL report (2007).
\bibitem{brown1}B. A Brown, N. J. Stone, J. R. Stone, I. S. Towner, and M. Hjorth-Jensen, Phys. Rev. C 71 (2005) 044317.
\bibitem{nazarewicz}W. Nazarewicz, P. Olanders, I. Ragnarsson, J. Dudek, G. A. Leander, P. Moller and E. Ruchowska, Nucl. Phys. A 429 (1984) 269.
\bibitem{cottle}P. D. Cottle, Phys. Rev. C 42 (1990) 1264.

\end{thebibliography}

\end{document}